\documentclass[12pt,titlepage,oneside]{article}

\usepackage{verbatim}
\usepackage{ametsocSS}

\linenumbers

\newcommand{\<}   {\langle}
\renewcommand{\>} {\rangle}

\newcommand{\ol}  {\overline}
\newcommand{\ca}  [1]{\mathcal{#1}}

\newcommand{\bs}  [1]{\boldsymbol{#1}}
\newcommand{\p}   {\partial}
\newcommand{\del} {\nabla}
\renewcommand{\vec} [1]{\boldsymbol{\mathrm{#1}}}   % boldface vectors
\newcommand{\mat} [1]{\mathsf{#1}}          % boldface sans serif matrices

\paperstatus{Submitted to JPO}

\title{A surface-aware projection basis for quasigeostrophic flow}

\author{K. Shafer Smith\thanks{\textit{Corresponding author address:}
         K. Shafer Smith, Courant Institute of Mathematical Sciences,
	 New York University, 251 Mercer St., New York, NY 10012. 
         \newline{E-mail: shafer@cims.nyu.edu}}}

\address{Center for Atmosphere Ocean Science\\ 
  Courant Institute of Mathematical Sciences\\
  New York University\\
  New York, NY}

\author{Jacques Vanneste}

\address{School of Mathematics\\
  University of Edinburgh\\ 
  Edinburgh, United Kingdom}

\begin{document}

\maketitle

\begin{abstract}
  Recent studies indicate that altimetric observations of the ocean's
  mesoscale eddy field reflect the combined influence of surface
  buoyancy and interior potential vorticity anomalies.  The former
  have a surface-trapped structure, while the latter have a more grave
  form.  To assess the relative importance of each contribution to the
  signal, it is useful to project the observed field onto a set of
  modes that separates their influence in a natural way.  However, the
  surface-trapped dynamics are not well-represented by standard
  baroclinic modes; moreover, they are dependent on horizontal scale.

  Here we derive a modal decomposition that results from the
  simultaneous diagonalization of the energy and a generalization of
  potential enstrophy that includes contributions from the surface
  buoyancy fields.  This approach yields a family of orthonomal bases
  that depend on two parameters: the standard baroclinic modes are
  recovered in a limiting case, while other choices provide modes that
  represent surface and interior dynamics in an efficient way.

  For constant stratification, these modes consist of symmetric and
  antisymmetric exponential modes that capture the surface dynamics,
  and a series of oscillating modes that represent the interior
  dynamics.  Motivated by the ocean, where shears are concentrated
  near the upper surface, we also consider the special case of a
  quiescent lower surface.  In this case, the interior modes are
  independent of wavenumber, and there is a single exponential surface
  mode that replaces the barotropic mode.  We demonstrate the use and
  effectiveness of these modes by projecting the energy in a set of
  simulations of baroclinic turbulence.

\end{abstract}

\clearpage

\section{Introduction}

Because direct observations of the ocean's interior are sparse,
satellite altimetry plays a crucial role in determining its
time-dependent, three-dimensional velocity structure.  This indirect
measurement process assumes that sea surface height variations are
dominated by currents with low-mode vertical structure, a result of
the stiffening action of rotation and ensuing barotropization.
Observations provide some support for this assumption, at least on
lateral scales of order the first internal deformation scale and
above.  For example, using currentmeter records in conjunction with
satellite obervations, \cite{Wunsch97} argues that the bulk of the
ocean's eddy kinetic energy resides in the barotropic and first
baroclinic modes.  In addition, a number of studies show a strong
correlation between the lateral size of eddies and the first internal
deformation scale \citep[e.g.][]{Stammer97,Chelton11}.

However, recent theoretical developments, supported by simulation and
improved analysis of satellite altimetry, suggest that surface signals
are not well-correlated with low-mode vertical structure, especially
for submesoscale motions.  In particular, \cite{Lapeyre06c} argue that
surface buoyancy and upper-ocean potential vorticity are
anti-correlated for eddying flow, and that the three-dimensional
velocity field may be obtained, assuming quasigeostrophy, from
knowledge of the surface buoyancy field alone.  The dynamics at the
upper surface in this view are closely related to the surface
quasigeostrophic (SQG) model \citep{Blumen82,Held95a}, and imply a
vertical structure with a surface-trapped component that is not well
represented by standard baroclinic modes.  This view is supported by
results from simulations \citep{LaCasce06,Klein08a}, as well as recent
analyses of satellite altimetry \citep[e.g.][]{Isern-Fontanet06,
  LeTraon08}.  Finally, in an atmospheric context, \cite{Tulloch09a}
have shown that lateral surface buoyancy gradients may interact with
interior mean potential vorticity gradients to excite baroclinically
unstable modes that generate SQG-like dynamics near the upper surface.
In simulations, the resulting kinetic energy spectrum near the surface
exhibits a steep $-3$ slope just below the deformation scale, and a
flatter $-5/3$ slope at smaller scales --- translated to the oceanic
context, this implies an energetic submesoscale dominated by the
surface mode.

One of the most widely used tools in oceanography is the projection of
the vertical structure of observed or simulated currents on simple
bases of functions. The above observations and modeling results lead
one to seek projection bases that faithfully represent both the
low-mode interior structure and the surface dynamics. The standard
basis of baroclinic modes, consisting of the eigenfunctions $\phi(z)$ of
the operator $\partial_z[f^2/N^2(z) ~\partial_z\phi ]$,
with homogenous boundary conditions $\partial_z\phi|_{z=0}=\partial_z \phi|_{z=-H} =
0$, fails in this respect.  By construction, it is a complete basis in
which to expand the streamfunction $\psi$ of flows provided they
satisfy the same homogeneous boundary conditions, which imply zero
surface and bottom buoyancy.  But for realistic flows with non-zero
surface buoyancy $b = f \p_z\psi|_{z=0}$, expansion in baroclinic
modes leads to a non-uniform convergence near $z=0$, and a very large
set of modes is required to capture the near-surface behaviour.

As noted by \cite{Lapeyre06c}, in quasigeostrophic theory, the
dynamical contribution of the surface buoyancy can be separated from
that of the interior potential vorticity: taking advantage of the
linearity of the inversion of the quasigeostrophic potential vorticity
(PV)
\begin{equation} \label{qgpv}
 q =\del^2\psi + \p_z\left(\frac{f^2}{N^2} ~\p_z\psi\right) 
\end{equation} 
the streamfunction may be decomposed into interior and surface parts,
$\psi = \psi^\text{int} + \psi^\text{surf}$ (assuming zero buoyancy at
the bottom), where $\psi^\text{int}$ satisfies (\ref{qgpv}) with
boundary condition $\partial_z \psi^\text{int}|_{z=0}=0$ while
$\psi^\text{surf}$ satisfies the zero-PV condition
$\del^2\psi^\text{surf} + \p_z(f^2/N^2 ~ \p_z\psi^\text{surf})=0$ with
$\p_z\psi^\text{surf}|_{z=0}=b/f$. The vertical structure of the
interior contribution can be expanded in the standard baroclinic
modes. By contrast, the surface contribution --- the only one retained
in SQG theory --- has a vertical structure determined by the zero PV
condition which couples horizontal and vertical dependence, reducing
to $\exp(\kappa N z/f)$, where $\kappa$ is the horizontal wavenumber,
in the case of constant $N$ and for $z \gg H$.

It is intuitively clear that an effective projection basis should
somehow combine modes similar to the baroclinic modes with modes that,
like the exponential modes of SQG theory, capture the dynamical
contribution of the surface buoyancy. A systematic method to obtain
such a basis has remained elusive, however. \cite{Tulloch09a} proposed
a heuristic model based on a barotropic and first baroclinic mode,
appended by exponential modes for each surface.  Similarly,
\cite{Lapeyre09} attempted to represent the full dynamics of the upper
ocean with a truncated set of standard baroclinic modes appended by an
exponential surface mode.  However, these hybrid modes do not
diagonalize the energy, since the surface and interior modes are not
orthogonal.  Moreover, because the surface modes depend on wavenumber
while the interior modes do not, the energetic overlap varies with
horizontal scale, increasing with increasing scale. These difficulties
stem from the fact that the addition of the exponential mode makes the
basis functions linearly dependent in a certain sense, leading to an
overcomplete frame rather than a basis. A consequence is that the
modal decomposition is non-unique. \cite{Lapeyre09} defined a unique
basis by requiring that it minimizes a certain functional, but the
results remained inconclusive.  An alternative basis, involving modes
satisfying the Dirichlet condition $\psi|_{z=0}=0$ together with the
barotropic mode, has recently been proposed by \citet{ScottRB12} but
this too suffers from a lack of orthogonality.

In this paper, we take a different approach and propose a new modal
basis (or rather a family of bases) that diagonalizes the energy and
effectively captures surface-intensified motion driven by
buoyancy. Our approach relies on the observation that there are
infinitely many possible (complete) bases onto which the flow may be
projected which diagonalize the energy.  As we show, a useful basis is
obtained by demanding that it simultaneously diagonalizes both the
energy and another quadratic invariant that generalizes potential
enstrophy to include the variances of the surface and bottom buoyancy
fields.  The relative weight of the potential enstrophy and buoyancy
variances in this invariant provide two parameters that determine the
basis uniquely. 

The eigenvalue problem that arises is similar to the standard vertical
mode problem, but retains a dependence on horizontal wavenumber, and
the eigenvalue appears in both the eigenvalue equation and its
boundary conditions.  In a limiting case, the standard baroclinic
modes are recovered --- for constant $N$ and $-H\le z\le 0$, these are
$\psi_n \propto \cos(n \pi z/H), \, n=0,\, 1,\ldots$. Another limiting
case, motivated by the ocean where shears are concentrated near the
upper surface but are weak at depth, leads to the simple basis
\begin{equation} \label{simbas}
\psi_0 \propto \cosh\left[N \kappa (z+H)/f\right], 
\quad \psi_n \propto \sin\left[(n-1/2) \pi z/H)\right],\, n=1,\,2,\ldots
\end{equation}
which includes the exponential mode of SQG theory. 

The paper is organized as follows.  In section 2 we construct a
generalized eigenvalue problem that defines the new basis.  In section
3, we derive analytical solutions and general results for two special
cases: constant $N$, for expository purposes, and an ocean-like case,
in which the lower boundary is assumed quiescent, leading to
(\ref{simbas}).  These modes are tested in section 4 on fields
generated from a set of high-resolution quasigeostrophic simulations
of baroclinic turbulence.  Finally, we discuss and conclude in section
5.

\section{Surface-aware basis}

Throughout the paper, we assume a horizontally-periodic domain bounded
vertically by rigid surfaces at $z=z^-$ and $z=z^+$, with total depth
$H=z^+-z^-$.  The horizontal periodicity allows us to Fourier
transform the equations in the horizontal plane, resulting in
separable dynamics and ordinary differential equations for the
vertical structure.  (In more general domains, the Fourier series can
be replaced by an expansion in eigenfunctions of the horizontal
Laplacian, and the results obtained here should hold essentially
unchanged.) The complex amplitudes of the quasigeostrophic potential
vorticity (PV) $q=q_{kl}(z)$, surface buoyancies (SBs) $b_{kl}^\pm$
and streamfunction $\psi=\psi_{kl}(z)$ are then related by
\begin{subequations}\label{pvb}
\begin{align}
\left(\frac{f^2}{N^2} \psi'\right)'-\kappa^2\psi &= q, ~~~ z^-<z<z^+ \label{pv}\\
\frac{f^2}{N^2H}\psi' &= b^\pm, ~~~ z=z^\pm,\label{b}
\end{align}
\end{subequations}
where $\kappa=(k^2+l^2)^{1/2}$ is the wavenumber magnitude, a prime
indicates a $z$ derivative, $f$ is the Coriolis frequency and $N=N(z)$
is the buoyancy frequency.  We include the non-standard factor $f^2/(N^2H)$
in our the definition of the SBs so that the SBs and PV have the same
dimension (inverse time), and because it ultimately yields a more
natural eigenvalue problem.  We have omitted the wavenumber subscript
on $q$, $b^\pm$ and $\psi$ and continue to do so onward, except where
confusion may occur.

The quasigeostrophic equation set has four quadratic invariants:
energy, potential enstrophy, and the buoyancy variance at each
surface.  At each wavenumber $\kappa$, these are
\begin{align*}
E_\kappa &= \frac{1}{2H}\int_{z^-}^{z^+} 
\left(\frac{f^2}{N^2}|\psi'|^2 + \kappa^2|\psi|^2 \right)~\mathrm{d}z\\
Z_\kappa &=  \frac{1}{2H}\int_{z^-}^{z^+} |q|^2~\mathrm{d}z\\
B_\kappa^\pm &= \frac{1}{2}|b^\pm|^2.
\end{align*}
Summing each quantity over $(k,l)$ gives the total invariant.

We seek to define a complete basis that diagonalizes the energy.  This
can be done in infinitely many ways. Our strategy is based on the
following principles: (i) we regard the energy as a functional, not of
the streamfunction, but of the PV and of the SBs; (ii) we exploit
standard results on the simultaneous diagonalization of quadratic
forms. Principle (i) is grounded in the quasigeostrophic model, which
makes it explicit that PV and SBs, taken together, make up the set of
dynamical variables.  Thus, the contribution of the SBs to the
dynamics is recognized; as a result, the bases we obtain naturally
represent data with non-zero surface buoyancies. Regarding (ii), we
recall a classical result from linear algebra: whereas there are
infinitely many bases diagonalizing a quadratic form
${\bs{x}}^\mathrm{T} \mat{A} \bs{x}$, where $\mat{A}$ is a symmetric
positive definite matrix, only one of these bases also diagonalizes
another quadratic form ${\bs{x}}^\mathrm{T} \mat{B} \bs{x}$
\citep[e.g.][]{HornJohnson90}. This is simply found by solving the
generalized eigenvalue problem $\mat{B} \bs{x} = \lambda \mat{A}
\bs{x}$. An analogous result applies to linear operators \citep[see,
e.g.][]{Goldstein80}.  Similarly, here we can define a unique basis by insisting
that it diagonalizes another quadratic form in addition to the energy
$E_\kappa$. A natural choice for this is a `generalized potential
enstrophy' that combines the remaining invariants into a single
quantity,
\begin{equation}\label{Pdef}
P_\kappa \equiv Z_\kappa + \alpha_+ B^+ + \alpha_- B^-
\end{equation}
where $\alpha_\pm > 0$ are (nondimensional) undetermined weights, the
choice of which will be discussed later. This approach yields a unique
basis for fixed $\alpha_\pm$.

To proceed, we require four objects: a vector structure that combines
the SBs and interior PV, an inner product that operates on this
vector, and two operators (analogous to the matrices $A$ and $B$
above) that give the energy and generalized potential enstrophy in
terms of the inner product. These are defined as follows:
\begin{description}
\item[Vector.] We define the `generalized potential vorticity
vector'\footnote{Notice that our $\bs{Q}$ bears a resemblance
  to the generalized potential vorticity of \cite{Bretherton66a},
  which in our notation is written
  \[
  Q_\text{B} = \left(\frac{f^2}{N^2} \psi'\right)'-\kappa^2\psi 
  - \frac{f^2}{N^2}\psi'\delta(z-z^+) + \frac{f^2}{N^2}\psi'\delta(z-z^-).
  \]
  Our notation makes it plain that the PV and SBs are independent, a
  point that the use of $Q_\text{B}$ might obscure.}
\begin{equation}\label{Qdef}
\bs{Q} \equiv \begin{pmatrix}
b^+\\
q(z)\\
b^-
\end{pmatrix}.
\end{equation}

\item [Inner product.] The specific choice of inner product is
  unimportant for the final results; we make what appears to be the
  simplest choice, namely
\begin{equation}\label{innerprod}
\<\bs{Q}_1,\bs{Q}_2\> 
=  \frac{1}{H}\int_{z^-}^{z^+}\bar{q}_1 q_2~\mathrm{d}z 
+ \bar{b}_1^+ b_2^+ + \bar{b}_1^-b_2^-,
\end{equation}
where the overbar denotes a complex conjugate.

\item[Operators.] With the definitions \eqref{Qdef} and
  \eqref{innerprod}, it is a simple matter to find the linear
  operators $\mathcal{E}$ and $\mathcal{P}$ such that
\begin{equation}\label{EPprod}
E_\kappa = \frac{1}{2}\<\bs{Q}, \mathcal{E} \bs{Q}\> \quad
\textrm{and} \quad 
P_\kappa = \frac{1}{2}\<\bs{Q}, \mathcal{P} \bs{Q}\>.
\end{equation}
These are given by
\begin{equation} \label{EPdef}
\ca{E} \bs{Q} = 
\begin{pmatrix}
\psi(z^+) \\
-\psi(z) \\ 
-\psi(z^-) 
\end{pmatrix}\quad
\textrm{and} \quad 
\ca{P} \bs{Q} = 
\begin{pmatrix}
\alpha_+ b^+ \\
q(z)\\ 
\alpha_- b^-
\end{pmatrix},
\end{equation}
where the streamfunction $\psi$ is the solution of \eqref{pvb}, given $q$
and $b^\pm$.  The first of these expressions is obtained after an
integration by parts; the second is immediate. These two operators are
positive definite and self-adjoint (see Appendix \ref{app:deriv} for
details).

\end{description}

The basis we seek is now given by the eigenfunctions $\bs{\xi}_n$ of
the generalized eigenvalue problem
\begin{equation}\label{eval}
\ca{P} \bs{\xi}_n = \mu_n^2 \ca{E} \bs{\xi}_n,
\end{equation}
where the eigenvalues $\mu_n^2$ are positive for all $n$.  To obtain
an explicit form for \eqref{eval}, we define the components of
$\bs{\xi}_n = [\xi_n^+, \xi_n(z), \xi_n^-]^\mathrm{T}$ analogous to
those of $\bs{Q}$, and the scalar streamfunctions $\phi_n(z)$ such
that $\ca{E}\bs{\xi}_n = [\phi_n(z^+), -\phi_n(z),
-\phi_n(z^-)]^\mathrm{T}$. In terms of these, the eigenvalue problem
reads
\begin{equation}\label{evalcomp}
\begin{pmatrix}
\alpha_+\xi_n^+\\   
\xi_n(z)\\ 
\alpha_-\xi_n^-
\end{pmatrix}
= \mu_n^2
\begin{pmatrix}
\phi_n(z^+) \\
-\phi_n(z)\\
-\phi_n(z^-) 
\end{pmatrix}.
\end{equation}
In view of \eqref{pvb}, this implies that the  $\phi_n$ satisfy
\begin{equation}\label{evalphi}
\left(\frac{f^2}{N^2} \phi_n'\right)' - \kappa^2 \phi_n = - \mu_n^2 \phi_n
~~~~\text{and}~~~~
\frac{f^2}{N^2H}\phi_n' = \pm\frac{\mu_n^2}{\alpha_\pm} \phi_n
~~\textrm{at} ~~ z=z^\pm.
\end{equation}
This eigenvalue problem is a key result of the paper. Its
eigenfunctions $\phi_n$, which are purely real, give the form of the
streamfunction corresponding to the basis eigenvectors
$\bs{\xi}_n$. The three components of these eigenvectors may be
derived from the $\phi_n$ using \eqref{evalcomp}, although, as shown
below, this is not necessary to project data onto the modes
$\bs{\xi}_n$.

By construction, the eigenfunctions are orthogonal for the products
$\< \cdot, \ca{E} \cdot \>$ and $\< \cdot, \ca{P} \cdot \>$. The
choice of normalization for the eigenvectors $\bs{\xi}_n$ is
inessential, but it is convenient to fix the energy of each mode to be
unity, that is, to take
\begin{equation} \label{ortho}
\< \bs{\xi}_m , \ca{E} \bs{\xi}_n \> = \frac{1}{H} \int_{z^-}^{z^+}
\left( \frac{f^2}{N^2} \phi_m' \phi_n' + \kappa^2 \phi_m \phi_n  \right) \, \mathrm{d}z 
= \delta_{mn}.
\end{equation}
The expression in terms of $\phi_m$ and $\phi_n$ is found by using
\eqref{evalcomp} and \eqref{evalphi} to eliminate $\bs{\xi}_m$,
$\bs{\xi}_n$ and the eigenvalues, then integrating by parts,
which removes boundary terms. Correspondingly,
\begin{equation} \label{xiPxiortho}
\< \bs{\xi}_m , \ca{P} \bs{\xi}_n \> 
= \frac{\mu_n^2}{H} \int_{z^-}^{z^+}
\left( \frac{f^2}{N^2} \phi_m' \phi_n' + \kappa^2 \phi_m \phi_n \right) \,\mathrm{d}z
=  \mu_n^2 \delta_{mn}
\end{equation}
and
\begin{equation}\label{PExiortho}
\< \mathcal{P}^{-1} \mathcal{E}\bs{\xi}_m , \mathcal{E} \bs{\xi}_n \>
= \frac{1}{H}\int_{z^-}^{z^+} \phi_m \phi_n~\mathrm{d}z 
+ \frac{\phi_m(z^+)\phi_n(z^+)}{\alpha_+} 
+ \frac{\phi_m(z^-)\phi_n(z^-)}{\alpha_-}
= \mu_n^{-2} \delta_{mn}.
\end{equation}
The latter relation \eqref{PExiortho} has the advantage of involving only the
undifferentiated streamfunctions, while the first relation
\eqref{ortho} is independent of the eigenvalues and $\alpha_\pm$.

The basis of eigenfunctions can be used to expand data: given $\bs{Q}$
or $\psi$, we can write
\begin{equation}\label{Qexp}
\bs{Q} = \sum_n a_n \bs{\xi}_n \quad \textrm{and} \quad 
\psi = \sum_n a_n \phi_n,
\end{equation}
where the $a_n$ are amplitude coefficients that can be found using one
of the orthogonality relations \eqref{ortho} or \eqref{xiPxiortho}; for
instance
\[
a_n = \<\bs{\xi}_n,\ca{E} \bs{Q} \> 
= \frac{1}{H} \int_{z_-}^{z_+} \left( \frac{f^2}{N^2} \phi_n' \psi' 
+ \kappa^2 \phi_n \psi \right) \, \mathrm{d} z.
\]
The energy and generalized potential enstrophy are then simply
\begin{equation}\label{EPdiag}
E_\kappa = \frac{1}{2}\sum_n  |a_n|^2 ~~~\text{and}~~~
P_\kappa = \frac{1}{2}\sum_n \mu_n^2|a_n|^2,
\end{equation}
respectively.

Note that, even though the eigenvalue problem \eqref{evalphi} is not
of the standard Sturm--Liouville form, because of the presence of the
eigenvalue $\mu_n^2$ in the boundary conditions, the basis of
eigenvectors can be shown to be complete in the sense that it provides
a representation of arbitrary vectors $\bs{Q}$ that converges as the
number of modes tends to $\infty$. This is discussed further in
Appendix \ref{app:deriv}.

Lastly, note that our choice of orthogonality conditions implies
slightly unfamiliar dimensions for the eigenfunctions.  Because
$[q],[b^\pm] \sim [T^{-1}]$ and $[\mu]\sim[L^{-1}]$ (where $T$ is time, $L$
is length, and braces mean ``dimensions of''), \eqref{eval} implies
that $[\bs{\xi}] \sim [L^{-2}] [\phi]$.  The orthogonality condition
\eqref{ortho} demands $[\phi] \sim [L]$ and therefore $[\bs{\xi}]\sim
[L^{-1}]$.    In the next section, the problem will be analyzed in an appropriate
nondimensional form.

\section{Structure of the surface-aware modes and special cases}

The approach described above provides a family of bases parameterized
by the values of $\alpha_+$ and $\alpha_-$. In principle, different
values can be chosen for different wavenumbers $\kappa$; here,
however, we restrict attention to choices of $\alpha_\pm$ that are
independent of $\kappa$. To clarify some general properties of the new
modes, we first recast the eigenvalue problem in non-dimensional form
with the substitutions $z \mapsto H z$, $\kappa \mapsto f/(N_0H) \,
\kappa$ and $\mu \mapsto f/(N_0H) \,\mu$, where $N_0$ is a typical
value of $N$;  thus the wavenumber and eigenvalue are
scaled by the approximate deformation length, $N_0H/f$.
The non-dimensional eigenvalue problem (\ref{evalphi}) then becomes
\begin{equation}\label{evalphinondim}
\left(s \phi_n'\right)' = -\lambda_n^2\phi_n
~~~~\text{and}~~~~
s \phi_n' = \pm\frac{\lambda_n^2+\kappa^2}{\alpha_\pm} \phi_n
~~\textrm{at} ~~ z=0, \, -1, ~~~\text{where}~~~s = \frac{N_0^2}{N^2(z)}
\end{equation}
and we have defined an alternative eigenvalue $\lambda_n$ such that
\begin{equation}\label{tlam}
 \mu_n^2 = \kappa^2+ \lambda_n^2.
\end{equation} 
Written in terms of $\lambda_n$, the eigenvalue equation takes the form of the standard
vertical mode equation, but with more complicated boundary conditions.

Analysis of the new eigenvalue problem \eqref{evalphinondim} is
complicated by its dependence on three independent parameters:
$\kappa$, $\alpha_+$ and $\alpha_-$.  Moreover, for each choice of
parameters, there is an infinite set of eigenvalues.  Since the
problem depends on the two weights $\alpha_\pm$ in a nearly equivalent
way, we proceed first by setting the weights equal and defining
$\alpha\equiv \alpha_+=\alpha_-$ (a case in which the weights differ
will be considered in a later subsection).  The nature of the
eigenproblem is then largely determined by the size of the boundary
condition coefficient $\mu_n^2/\alpha$: when $\mu_n^2/\alpha
\rightarrow 0$, the boundary conditions revert to the standard case
$\phi_n' = 0$ at the top and bottom, while when $\mu_n^2/\alpha
\rightarrow \infty$, the boundary conditions become $\phi_n = 0$ at
the top and bottom.  However, more subtle possibilities arise as well,
because unlike the standard vertical mode problem, $\lambda_n$ may be
imaginary (although $\mu_n$ is always real).  When $\lambda_n$ is
real, the modes are oscillatory, but when it is imaginary, the modes
are evanescent --- these can be interpreted either as surface modes or
as extensions of the barotropic mode.

This interpretation is suggested by examining the eigenvalue problem
in two limiting regimes:
\begin{description}
\item[$\kappa^2 \ll \alpha$:] modes with real $\lambda$ satisfy the
  simplified boundary condition $(s \phi_n')=\pm \lambda^2_n
  \phi_n/\alpha$ at $z=0,\, -1$ which further reduces to $\phi_n'=0$
  for $\alpha \gg 1$, corresponding to the standard baroclinic
  modes.\footnote{This approximation is not uniform in $n$ but breaks
    down for highly oscillatory modes, with $\lambda_n=O(\alpha)$,
    which satisfy $\phi'=O(\alpha)\not=0$ at $z=0,\, -1$ and thus
    differ from the standard high-$n$ baroclinic modes.}  These are
  complemented by a barotropic mode for which the first approximation
  $\lambda=0$ can be refined to the purely imaginary $\lambda = i
  \kappa \sqrt{2/\alpha}$.

\item[$\kappa^2 \gg \alpha$.] In this case, almost all modes have
  $\mu_n^2= \kappa^2 + \lambda_n^2 \gg \alpha$ and hence satisfy the
  simplified boundary conditions $\phi_n=0$ at $z=0,\, -1$. There are
  two additional modes, however, for which $\mu_n^2 = O(\alpha)$ and
  hence $\lambda \sim i \kappa$. These solve
\begin{equation}\label{surflim}
  \left(s \phi_n'\right)'  -\kappa^2\phi_n \simeq 0
  ~~~~\text{with}~~~~
  s \phi_n' = \pm\frac{\mu_n^2}{\alpha_\pm} \phi_n
  ~~\textrm{at} ~~ z=0, \, -1,
\end{equation}
and can be recognized as surface modes, with zero interior PV.

\end{description}

\subsection{Analytical solutions for constant $N$}

\begin{figure}
\begin{center}
\vskip 1cm
\includegraphics[width=.7\textwidth]{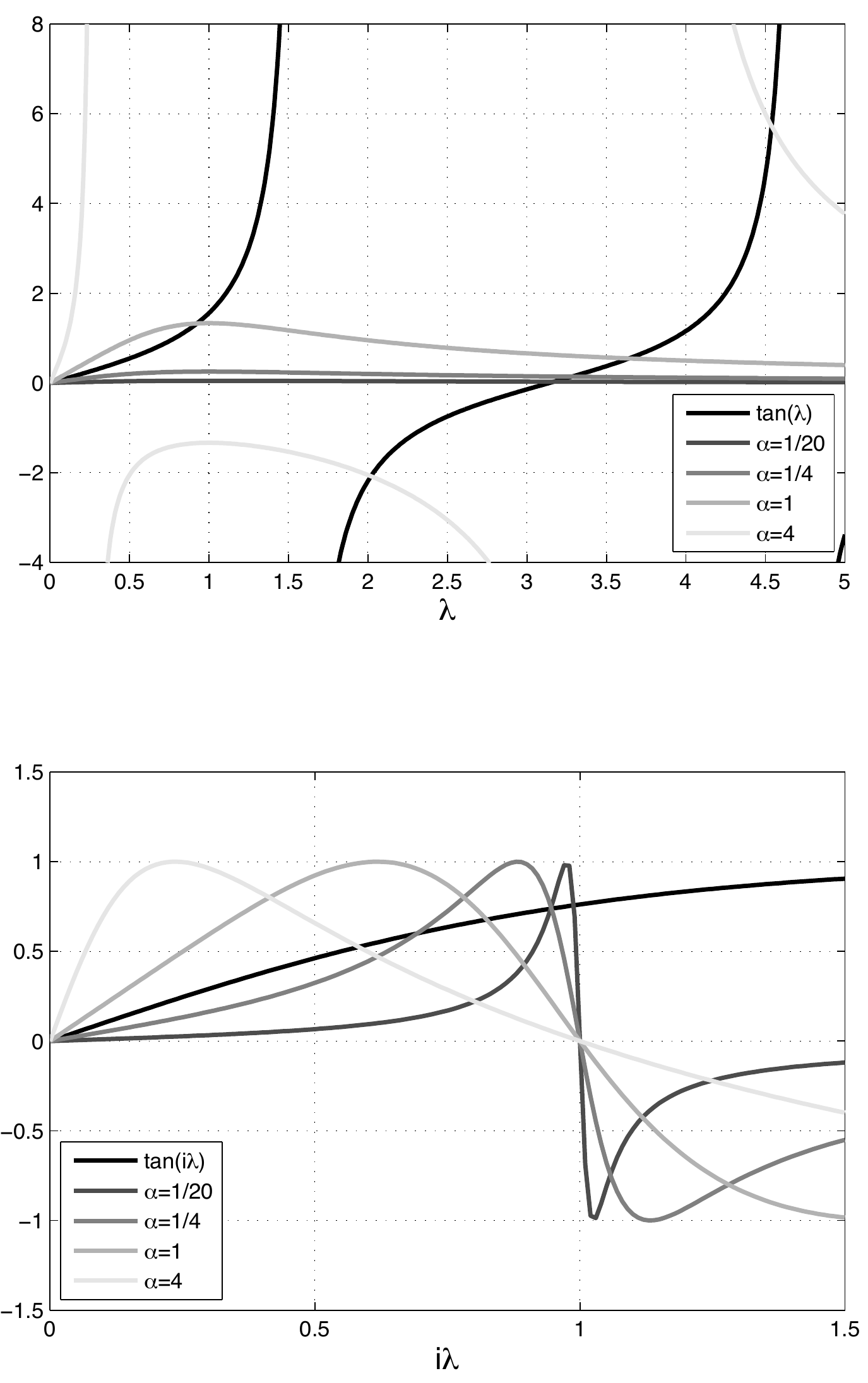} 
\caption{Graphical solutions for eigenvalues with constant $N$ for
  $\kappa =1$.  The left panel shows the left and right hand sides of
  Eq. \eqref{alg1}, and the right panel shows Eq. \eqref{alg2}.}
\label{tanfg}
\end{center}
\end{figure}

In the special case of constant stratification, or $s=1$, the 
eigenvalue problem \eqref{evalphinondim} can be solved in closed form.  Writing
the solutions as
\[
\phi_n = A  \cos(\lambda_n z) + B \sin(\lambda_n z), 
\]
where $A$ and $B$ are integration constants,
and imposing the boundary conditions leads to an algebraic equation
for $\lambda_n$, which may be either real or imaginary.  For
$\lambda_n^2>0$, the characteristic equation (dropping the subscript $n$) is
\begin{equation} \label{alg1}
  \tan \lambda = \frac{(\alpha_++ \alpha_-) \lambda
    (\lambda^2+\kappa^2)}{(\lambda^2+\kappa^2)^2 - \alpha_+\alpha_- \lambda^2}.
\end{equation}
For $\lambda^2<0$ we define $\tilde{\lambda}=i \lambda$ and obtain
\begin{equation} \label{alg2} \tanh \tilde \lambda =
  \frac{(\alpha_++\alpha_-) \tilde{\lambda}
    (\kappa^2-\tilde{\lambda}^2)}{(\kappa^2-\tilde{\lambda}^2)^2 + \alpha_+\alpha_-
    \tilde{\lambda}^2}.
\end{equation}
Equations \eqref{alg1} and \eqref{alg2} are suitable for a graphical
analysis.  Fig.~\ref{tanfg} shows that there are infinitely many
solutions to \eqref{alg1} (top panel) and one or two solutions to
\eqref{alg2} depending on $\alpha_\pm$ (bottom panel; in both cases we
set $\alpha \equiv \alpha_+ = \alpha_-$). An important parameter is
the ratio of the slopes of the right- and left-hand sides of
\eqref{alg1} and \eqref{alg2} at $\lambda=0$, which in both cases is 
\[
\frac{\alpha_++ \alpha_-}{\kappa^2} \equiv \tilde\kappa^{-2}
\]
When $\tilde\kappa<1$ there is only one solution to \eqref{alg2},
and there is a solution of \eqref{alg1} with $\lambda < \pi/2$.  On
the other hand, if $\tilde\kappa > 1$, there are two solutions to
\eqref{alg2} (note that the maximum of the right-hand side of
\eqref{alg2} is $1$), and there may or may not be a solution of
\eqref{alg1} for $\lambda<\pi/2$.\footnote{Note also that if
  $\alpha_+\alpha_->4\kappa^2$, the denominator of the right-hand side
  of \eqref{alg1} goes to $0$, but stays finite otherwise: the
  existence of a 0 in the denominator determines whether there is a
  solution to \eqref{alg1} with $\lambda<\pi/2$ in the case $\tilde\kappa^{-2}>1$.}

\begin{figure}
\begin{center}
\vskip 1cm
\includegraphics[width=.9\textwidth]{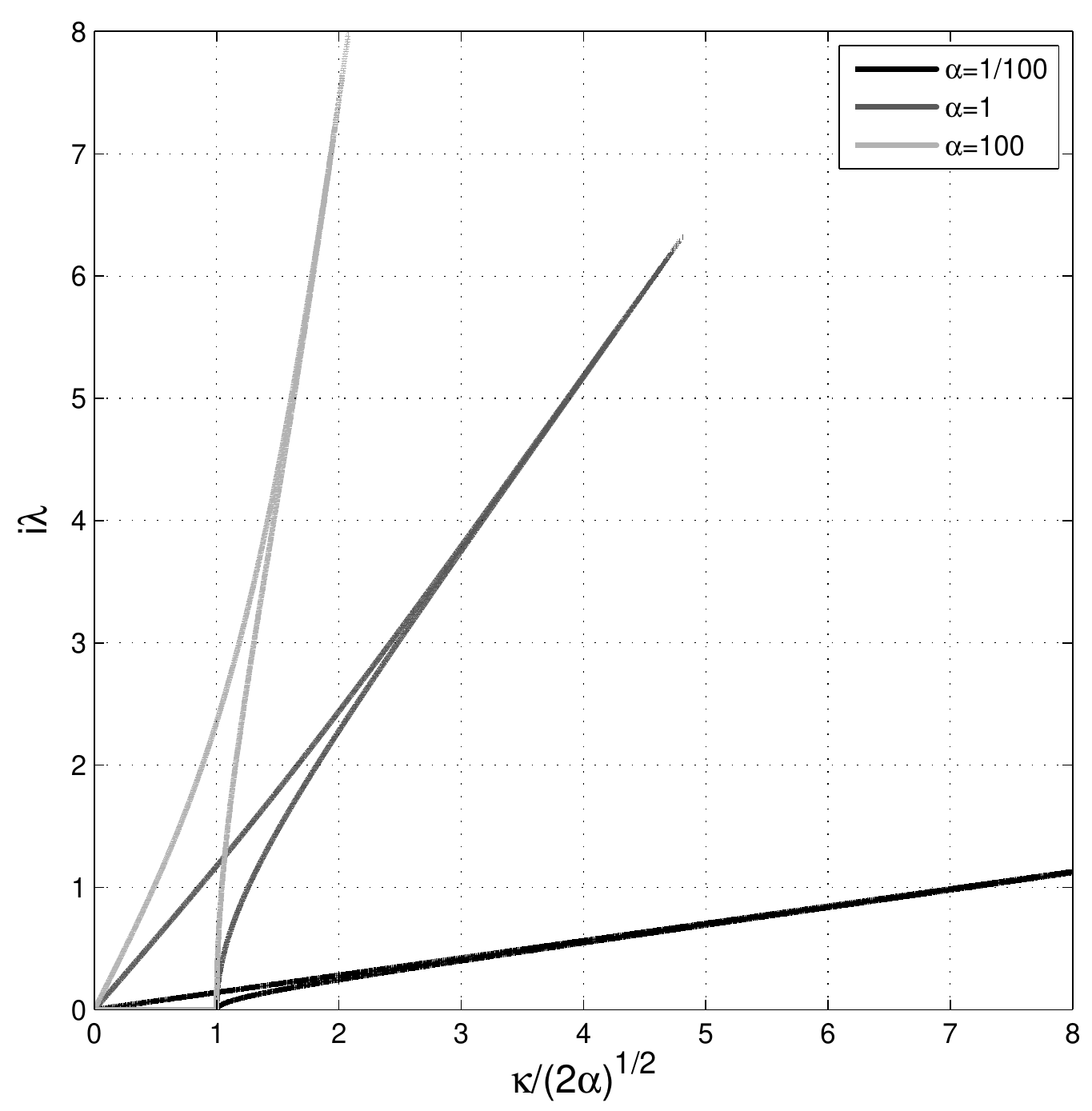}
\caption{Solutions to \eqref{alg2}, with $\kappa$ scaled by
  $\sqrt{2\alpha}$, the cutoff separating cases with one or two
  solutions for imaginary $\lambda$.}
\label{nutvsk}
\end{center}
\end{figure}

\begin{figure}
\begin{center}
\vskip 1cm
\includegraphics[width=.9\textwidth]{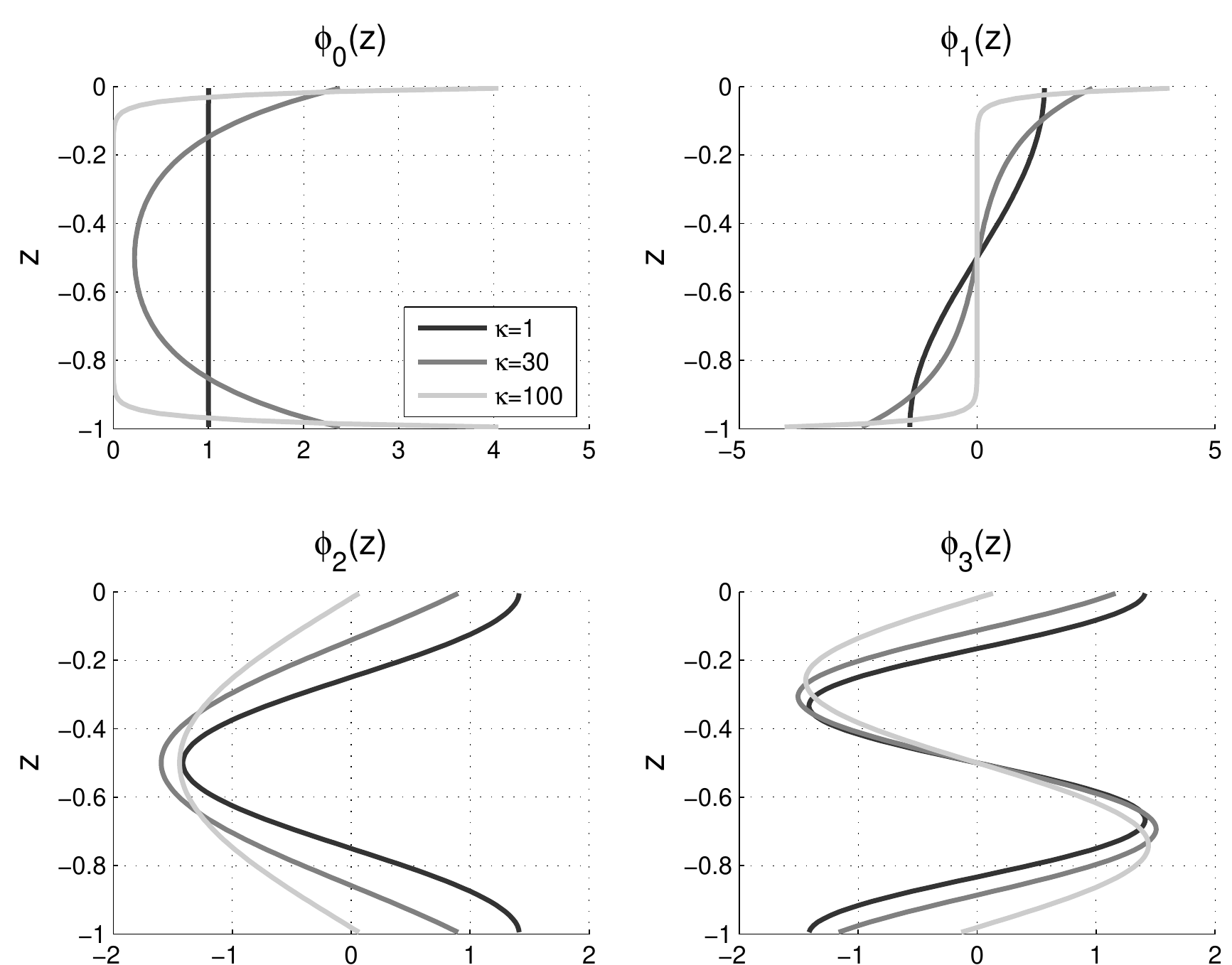}
\caption{The first four eigenfunctions $\phi_n$ for the constant-$N$
  case, with $\alpha_+=\alpha_-=100$ and $\kappa=1, 30, 100$.}
\label{modes1to4constN}
\end{center}
\end{figure}

The solution to \eqref{alg2} gives either a generalization of the
barotropic mode, in the case of a single solution, or two modes that
capture the vertical structure of the surface modes.  Setting $\alpha
\equiv \alpha_+ = \alpha_-$, these solutions are plotted as functions
of $\tilde\kappa$ in Fig. \ref{nutvsk}: there are two solutions when
$\tilde\kappa >1$, but only one otherwise. The limiting solutions
discussed in the previous section can be derived explicitly. In the
limit $\tilde\kappa^2 = \kappa^2/(2\alpha) \ll 1$, the single solution
of \eqref{alg2} is given by $\tilde\lambda \sim
\kappa\sqrt{2/\alpha}$, with eigenfunction $\phi \propto 1$, which can
be interpreted as the barotropic mode.  For $\tilde\kappa^2\gg 1$, the
two solutions can be identified as surface intensified modes, one
symmetric and the other antisymmetric about the center of the domain,
explicitly given by
\[
\phi_0 \propto \cosh\left[\kappa(z+\tfrac{1}{2})\right] 
~~~\text{and}~~~
\phi_1\propto \sinh\left[\kappa(z+\tfrac{1}{2})\right],
\]  
with eigenvalues $\mu_0/\alpha = \kappa\tanh\kappa$ and $\mu_1/\alpha
= \kappa\coth\kappa$.  For $\kappa\gg 1$, the eigenvalues are nearly
identical, so that linear combinations of the eigenfunctions will also
satisfy the eigenvalue problem --- in particular, one can construct
separate upper-surface and lower-surface modes.  For real $\lambda$,
the right-hand side of \eqref{alg1} tends to zero for both large and
small $\kappa$, leading to eigenvalues $\lambda_n = n \pi, \,
n=1,2\ldots$ The eigenfunctions, however, differ in the two cases: for
$\tilde\kappa \ll 1$, they have the standard form $\phi_n
\propto \cos(n\pi z)$, but for $\tilde\kappa \gg 1$, they are
$\phi_n \propto \sin(n\pi z)$.  The first four modes, for $\alpha=1$
and a range of $\kappa$ are plotted in Fig.~\ref{modes1to4constN}.

\subsection{An oceanic special case} \label{sec:ocean}

Here we consider a case that is potentially the most relevant to the ocean, where
shears near the surface may lead to surface-intensified modes, while
the quiescent abyss may be more naturally represented by the standard
boundary condition, $\phi' = 0$ at the bottom.  The relevant limits
for this case are $\alpha_+\ll 1$ and $\alpha_-\rightarrow\infty$, in
which case the eigenvalue problem reduces to
\begin{subequations}\label{oceig}
\begin{align} 
(s \phi_n')' &= - \lambda_n^2 \phi_n, \quad \textrm{with} \quad
 \phi_n|_{z=0} = 0,  \quad\quad\quad  \phi_n'|_{z=-1}=0, \label{oceiga} \\
(s \phi_0')' - \kappa^2 \phi_0 &= 0, \quad \quad \quad \textrm{with} \quad 
s \phi'_0|_{z=0} = \frac{\mu_0^2}{\alpha_+} \phi_0 , \quad \phi_0'|_{z=-1}=0.\label{oceigb}
\end{align}
\end{subequations}
to leading order in $\alpha_+$.  The solutions $\phi_n, \,
n=1,2\ldots$ to (\ref{oceiga}) describe interior modes, while $\phi_0$
is the solution to (\ref{oceigb}) with $\mu_0^2 / \alpha_+ = O(1)$ and
represents a zero PV, surface-intensified mode.

Note that the structure of the interior modes, like that of the
standard baroclinic modes, is independent of $\kappa$; the
normalization of the mode energy that we have chosen however leads to
$\kappa$-dependent normalization factors. Since we concentrate on the
leading-order approximation to the eigenvalue problem as $\alpha_+ \to
0$, all the modes, including the surface-intensified one, are
independent of $\alpha_+$ and so are the normalisation factors
(because the energy does not involve $\alpha_+$). Only the eigenvalue
$\mu_0^2$ depends (linearly) on $\alpha_+$, although the approximation
$\mu_0^2=0$ can be made to conclude, in particular, that the
surface-intensified mode has a generalized enstrophy which vanishes to
leading order.

Recently, \citet{ScottRB12} proposed to use the eigenfunctions of
(\ref{oceiga}), forming what they term a Dirichet basis, in
conjunction with the barotropic mode. While this set of functions,
like that obtained by adding a surface mode to the standard baroclinic
basis \citep{Lapeyre09}, does not diagonalize the energy, it is
remarkable that this is achieved by the complete set of solutions of
(\ref{oceiga}) and (\ref{oceigb}), that is, by the Dirichlet basis
plus a surface mode.

%Note the following properties of these `ocean'-modes:
%\begin{itemize}
%\item Because $\alpha_+$ is taken to be 0 for modes with $n>0$, but
%$O(\mu_0^2)$ for $n=0$, the surface mode $\phi_0$ is perfectly
%orthogonal to the interior modes $\phi_n$ under the
%$\alpha_+$-independent condition \eqref{ortho}, but only
%approximately orthogonal under condition \eqref{PExiortho}.
%\item Apart from the eigenvalue for the surface mode, the ocean basis
%  is independent of $\alpha_+$. 
%\item Except for the normalization factor, the interior modes in this
%  special case are independent of $\kappa$.
%\end{itemize}

For constant $N$ (or $s=1$), the solutions to \eqref{oceig} may be computed
explicitly;  they are
\begin{subequations}\label{oceigsol}
\begin{align}
\phi_0 &= A\cosh\left[\kappa(z+1)\right], \qquad 
A \equiv \sqrt{\frac{2}{\kappa \sinh (2 \kappa)}} \\
\phi_n &= B \sin \left[\left(n-\frac{1}{2}\right)\pi z\right], \quad
B \equiv \sqrt{\frac{2}{\pi^2(n-1/2)^2+\kappa^2}} 
\end{align}
\end{subequations}
with eigenvalues $\mu_0^2 = \alpha_+ \kappa \tanh \kappa$
(corresponding to $\tilde\lambda \simeq
\kappa-(\alpha_+/2)\tanh\kappa$) and $\lambda_n = (n-1/2)\pi$ with
$n=1,2 \ldots$.  Their dimensional form was given by (\ref{simbas}) in
the introduction. Again, note that the dependence on $\kappa$ of the
coefficient for the interior modes is due to the normalization choice,
but is irrelevant for the projection of data.

\section{Use of new basis for the projection of simulated data}

As a demonstration, we use the new basis to project the energy in
three simulated turbulent flows, each generated by baroclinic
instability of a fixed mean state in a horizontally-periodic
quasigeostrophic model.  The numerical model is spectral in the
horizontal, and finite-difference in the vertical --- it is the same
as used in, for example, \citet{SmithFerrari09}. Energy is dissipated
by linear bottom drag, and enstrophy is removed by a highly
scale-selective exponential cutoff filter \citep{Smith02b}.  In all
cases, the model resolution is $512 \times 512 \times 100$.

We analyze results from three simulations.  These first two are based
on highly idealized flows, and will be used to demonstrate the
fundamental structure of the basis, and how the partition of energy
depends on both the nature of the flow, and on the choice of the
nondimensional weights $\alpha_\pm$.  The third simulation is based on
a more realistic, ocean-like mean state, and is designed to explore
the oceanic special case considered at the end of the last section.
To project the simulated data onto the new basis, one must consider
the generalized matrix eigenvalue problem that results from the
particular vertical discretization used in the model.  The details of
the construction of the basis in this discretization are given
explicitly Appendix B.

\subsection{Idealized `interior' and `surface' baroclinic instability
  simulations}

Both idealized flows have constant stratification $s=1$, a ratio of
domain scale to deformation scale equal to $4$ and $\beta=0$, but mean
states that generate different types of baroclinic instability.  The
first simulation, is forced by an `interior instability,' with a mean
flow that projects onto the first (standard) baroclinic mode, $U(z) =
\cos \pi z$.  Flows of this type are unstable due to a sign change of
the mean interior PV gradient, but have no mean SB gradients, since
$B^\pm_y \propto U_z|_{z=0,-1}=0$ --- we refer to this simulation as
BC1.  The second flow is forced by an Eady mean state, with a linear
mean shear $U(z) = z$, so the instability is driven by mean SB
gradients $B^\pm_y = 1$, resulting in energy generation near the two
surfaces.

\begin{figure}
\begin{center}
\vskip 1cm
\includegraphics[width=.8\textwidth]{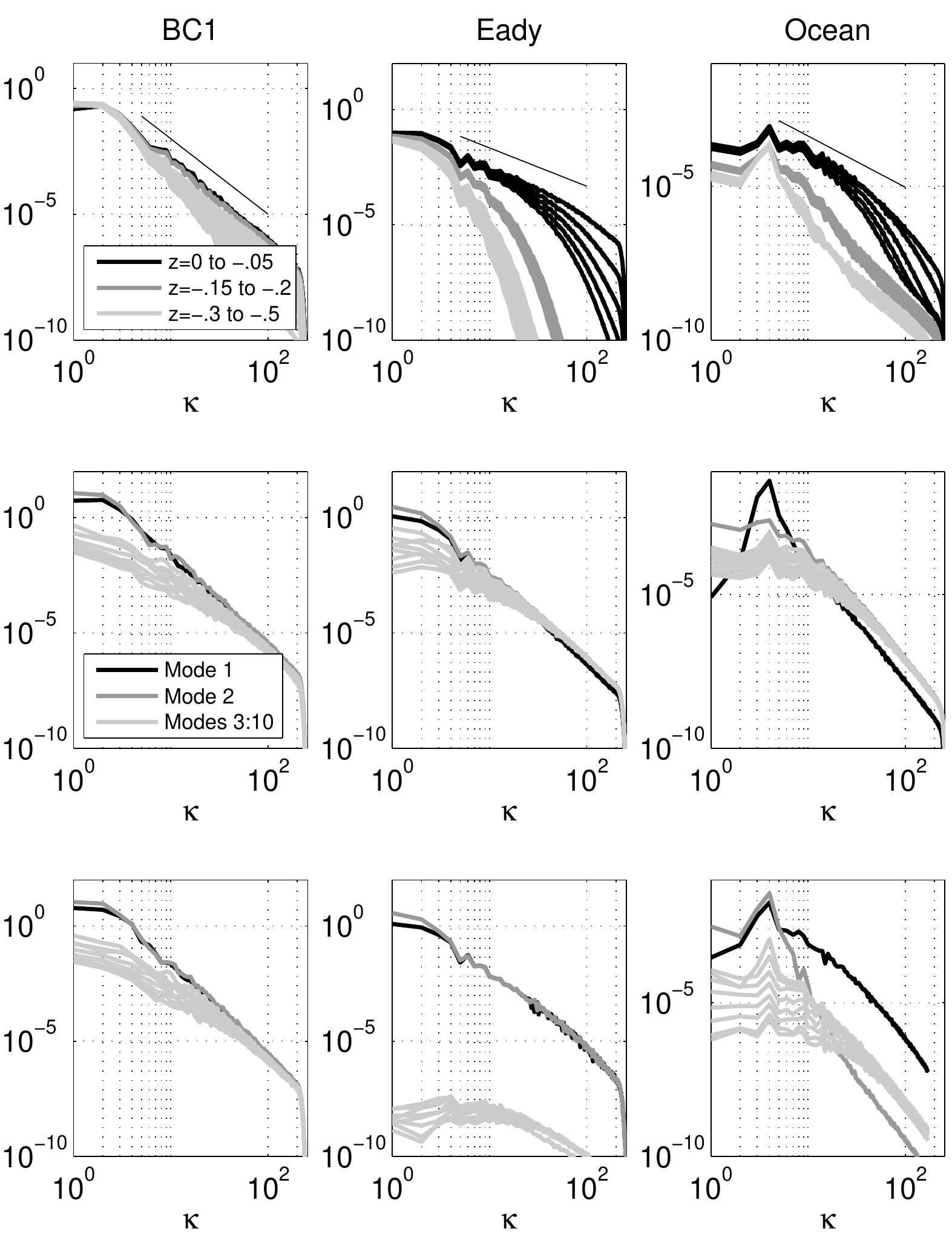}
\caption{Energy spectra for the BC1 (left), Eady (middle) and Ocean (right)
  simulations. Top panels: spectra for selected vertical levels (see legend).  Middle:
  spectra from fields projected onto standard vertical modes (modes 1,
  2 and 3--10 are shown).  Bottom:
  spectra from fields projected onto new modes, with
  $\alpha_+=\alpha_-=10^6$ for the BC1 case,
  $\alpha_+=\alpha_-=10^{-4}$ for the Eady case and 
  $\alpha_+=2$, $\alpha_-=10^6$ for the Ocean case. } 
\label{SIspec}
\end{center}
\end{figure}

The simulations are run to statistically steady state, and snapshots of the
steady-state prognostic fields of each are used to compute horizontal
(total) energy spectra.  The upper panels of Fig. \ref{SIspec},
display the horizontal spectra for the BC1 (left) and Eady (middle) simulations for a
few vertical levels $z$ (the right-hand column plots will be discussed
in the next subsection).  It is immediately apparent that the energy
in the BC1 simulation is spread rather evenly over depth; by
contrast, the energy in the Eady simultion is largly concentrated at
the two surfaces.  The panels in the middle row of Fig. \ref{SIspec} show the
first few modes of the energy projected onto the standard basis,
$\phi_n(z) \propto \cos(n\pi z)$, $n=1, 2, ...$ (the baroclinic modes)
and $\phi_0 \propto 1$ (the barotropic mode).  Consistent with the
$z$-dependence of the energy in the upper panel, the energy in BC1 is
largely captured by the barotropic and first baroclinic modes.  By
contrast, the energy in the Eady case seems to be distributed evenly
across the barotropic and a large number of baroclinic modes,
effectively demonstrating the failure of the standard modes to provide
any insight into the energy partition in a case with large energy
near the surfaces.  

The bottom panels of Fig. \ref{SIspec} display the energy spectra for
the first few modes in the projection onto the new basis (BC1, left
panel; Eady, middle panel).  Anticipating that the BC1 simulation is
best represented by the standard baroclinic basis (recovered from the
generalized basis in the limit $\alpha_\pm \gg 1$), while the Eady
simulation is best represented on the generalized basis in the limit
$\alpha_\pm\ll 1$, we chose $\alpha_\pm = 10^6$ for the former and
$\alpha_\pm = 10^{-4}$ for the latter.  As is apparent, the
generalized basis with the appropriate weights more efficiently
captures the surface energy in the Eady simulation much better than
the standard basis.

To quantify the choice of $\alpha_\pm$, we consider the projection of
energy in both the BC1 and Eady simulations with the generalized basis
using weights ranging from $\alpha_\pm = 10^{-3}$ to $10^3$ (always
holding $\alpha = \alpha_+=\alpha_-$) and ask, for what weights is the
energy captured by the least number of modes?  A simple diagnostic for
this, the ratio of the energy contained in the first two modes to the
total energy as a function of $\alpha$, is shown in in
Fig.~\ref{maxEmode}.  The results indicate that extreme values of
$\alpha$ are best suited for the BC1 ($\alpha \to \infty)$ and Eady
($\alpha \to 0$) simulations, thus confirming our choice for
Fig.~\ref{SIspec}. In the next section we examine a third simulation
where the interior and surface contributions are more balanced, so
that intermediate values of $\alpha_\pm$ may be expected to be
relevant.
 
\subsection{A semi-realistic oceanic simulation}

\begin{figure}
\begin{center}
\vskip 1cm
\includegraphics[width=\textwidth]{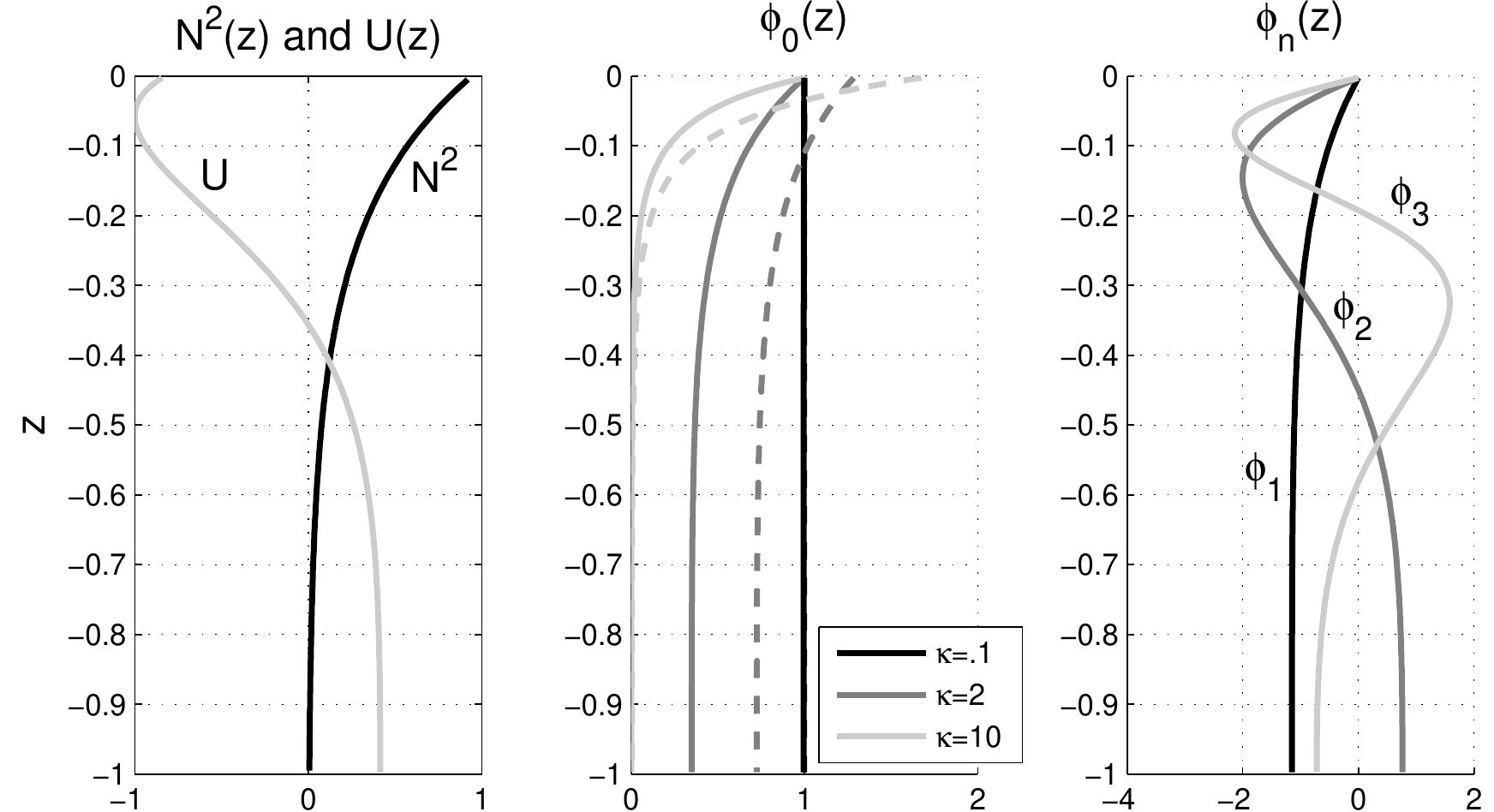}
\caption{Left: $N^2(z)$ and $U(z)$ for the Ocean simulation.
  Middle: the surface mode $\phi_0(z)$ with
  $\alpha_-\rightarrow\infty$ and $\alpha_+\ll 1$ (solid) and
  $\alpha_+=2$ (dashed), for a range of wavenumbers $\kappa$ (see
  legend).  The $\kappa=.1$ lines are on top of each other.  Right:
  The first three interior modes with $\alpha_+\ll 1$ and
  $\alpha_-\rightarrow\infty$.  }
\label{ocemodes}
\end{center}
\end{figure}

\begin{figure}
\begin{center}
\vskip 1cm
\includegraphics[width=1\textwidth]{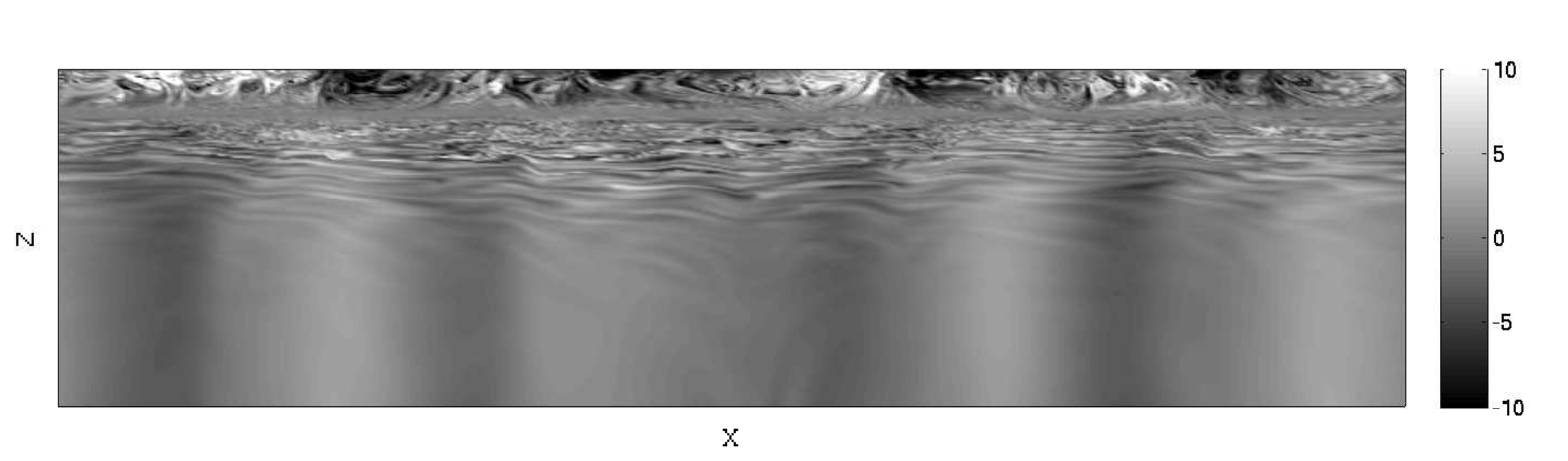}
\caption{Vertical slice of PV snapshot from the Ocean simulation.  The
  flow has a complicated structure in the upper ocean, masking a more
  uniform flow at depth. }
\label{pvslice}
\end{center}
\end{figure}

The third simulation is driven by a mean state typical of the
mid-latitude ocean.  It uses an exponential mean stratification $N^2 =
N_0^2 \exp(z/h)$, so that $s=\exp(-z/h)$, with $h=0.2$, intended to
represent the pycnocline.  The mean shear is
$U(z)=h(z+1-h)\exp(z/h)+g(z)+C$, where $g(z)$ is the first standard
baroclinic eigenfunction of the operator $(sg')'=-\lambda^2g$, with
$g'=0$ at $z=0,-1$, so that $U$ is surface-intensified with $U'(0)=1$
and $U'(-1)=0$.  The constant $C$ is set to ensure
$\int_{-1}^0U(z)~dz=0$. Both $U(z)$ and $N(z)$ are plotted in the
top panel of Fig.~\ref{ocemodes}.  Note that $U$ is baroclinically
unstable due to both an internal sign change of the mean PV
gradient, and to the interaction of the mean interior PV gradient
$Q_y$ with the mean upper SB gradient $B^+_y$.  Consistent with the
assumptions of the ocean modes, the lower SB gradient $B^-_y = 0$. The
ratio of the domain scale to the first baroclinic deformation
radius (as determined by $\lambda^{-1}$) is $5$. The nondimensional
Coriolis gradient $\beta U_0L_D^{-2}=1.2$, and energy is dissipated by
a linear drag $rL_dU_0^{-1}=0.4$.  The steady-state turbulent flow has
a complicated vertical structure, as evidenced by the vertical slice
of the PV shown in Fig.~\ref{pvslice}.

The energy spectra for the flow are shown in the right panels of
Fig.~\ref{SIspec}, just as for the BC1 and Eady cases.  The energy
spectra by vertical level again indicates a very surface-intensified
flow, but this time, the flow falls off from a $-5/3$ spectral slope
to a more energetic interior than was the case for the Eady
simulation.  Projection onto the standard vertical modes (middle right
panel) indicates a peak in the barotropic mode, but otherwise energy
is spread evenly over a large number of baroclinic modes.  Projection
onto a generalized basis is shown in the bottom right panel.  For this
simulation with no buoyancy activity at the bottom, it is natural to
use a basis with $\alpha_- \to \infty$. The maximum in the
ratio of the energy in modes 1 and 2 to total energy shown in
Fig.~\ref{maxEmode} suggests that the value $\alpha=\alpha_+=2$ is
appropriate. The first few modes of the corresponding basis are shown
in the bottom panels of Fig~\ref{ocemodes}. This
is the basis chosen for Fig.~\ref{SIspec}, and indicates that the
projection is very effective, with most of the energy captured by the
surface and modified first baroclinic modes.  An alternative basis is
the `oceanic' basis of section \ref{sec:ocean} which takes $\alpha_+
\ll 1$. The spectra obtained with this basis (not shown) are
essentially identical to those obtained for $\alpha_+=2$. This
suggests that the results are insensitive to the precise value of
$\alpha_+$ and that `oceanic' basis may be a good default choice to
analyse typical ocean data.

\section{Conclusion}

This paper presents a family of basis functions designed for the
projection of three-dimensional ocean velocity data.  The bases
diagonalize both the quasigeostrophic energy and a generalization of
the quasigeostrophic potential enstrophy that includes contributions
from the buoyancy variances at the upper and lower surfaces.  The
family of bases is parameterized by the weights $\alpha_\pm$ assigned
to the surface buoyancy variances --- the standard baroclinic modes
are recovered in the limit $\alpha_\pm\rightarrow\infty$, but the
modes obtained in the opposite limit allow for efficient
representation of the surface buoyancy variances. The bases should
prove advantageous in a number of applications, from projection of
observations to the derivation of highly truncated theoretical
models. Their main drawback compared to the standard basis of
baroclinic modes is the dependence of the modes on the wavenumber
$\kappa$ which implies a lack of separation between the horizontal
vertical structure in physical space. This drawback is unavoidable if
some of the modes are to reflect the SQG contribution; it is minimised
for the `oceanic' basis obtained for $\alpha_+ \to 0, \, \alpha_- \to
\infty$ since all but one modes have a $\kappa$-independent structure.

\begin{figure}
\begin{center}
\vskip 1cm
\includegraphics[width=\textwidth]{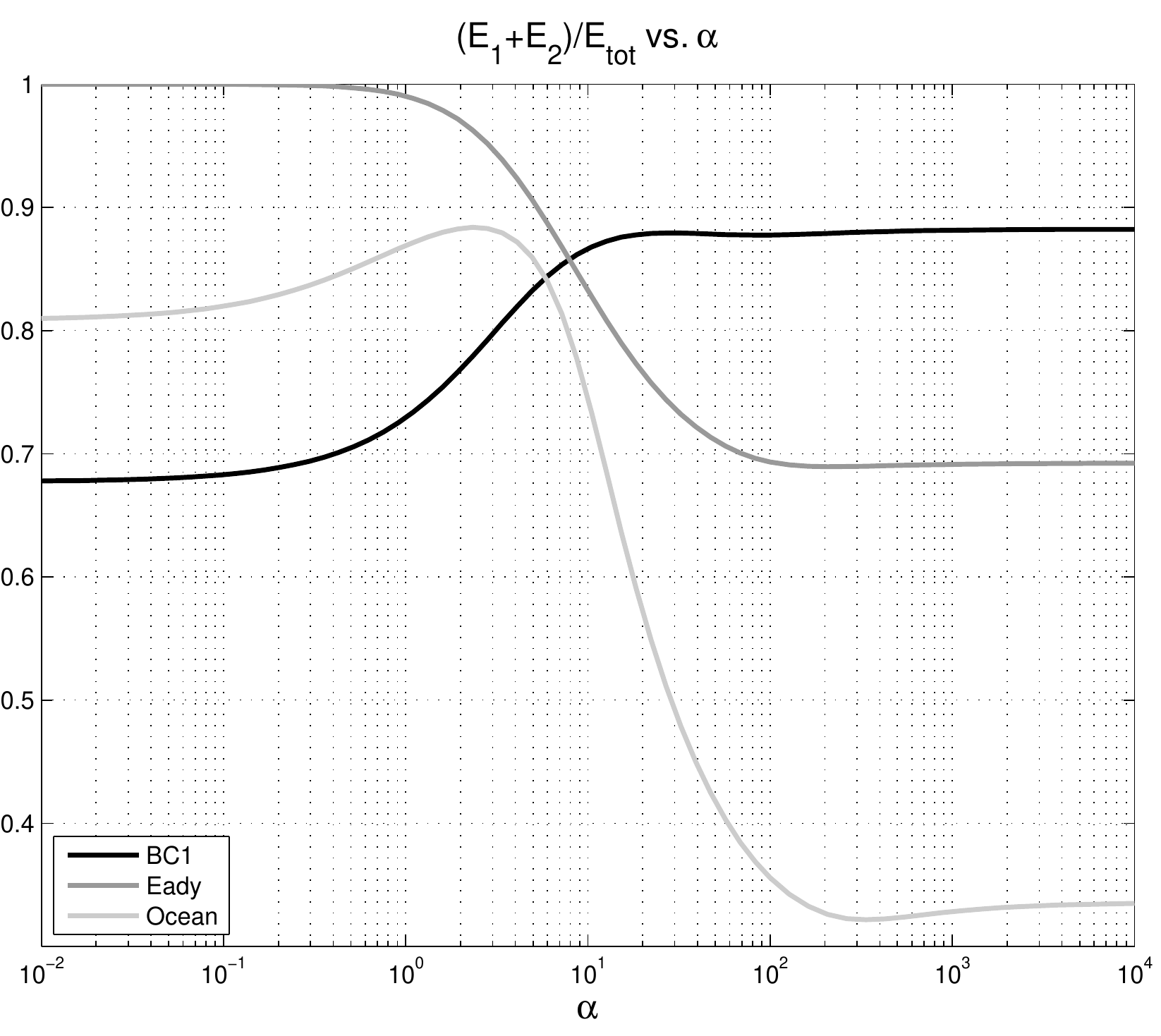}
\caption{Ratio of the energy content of the first two modes to the
  total energy as a function of $\alpha=\alpha_+=\alpha_-$ for the BC1
  and Eady simulations, and as a function of $\alpha=\alpha^+$ (with
  $\alpha_- \to \infty$) for the Ocean simulation.}
\label{maxEmode}
\end{center}
\end{figure}

\begin{figure}
\begin{center}
\includegraphics[width=\textwidth]{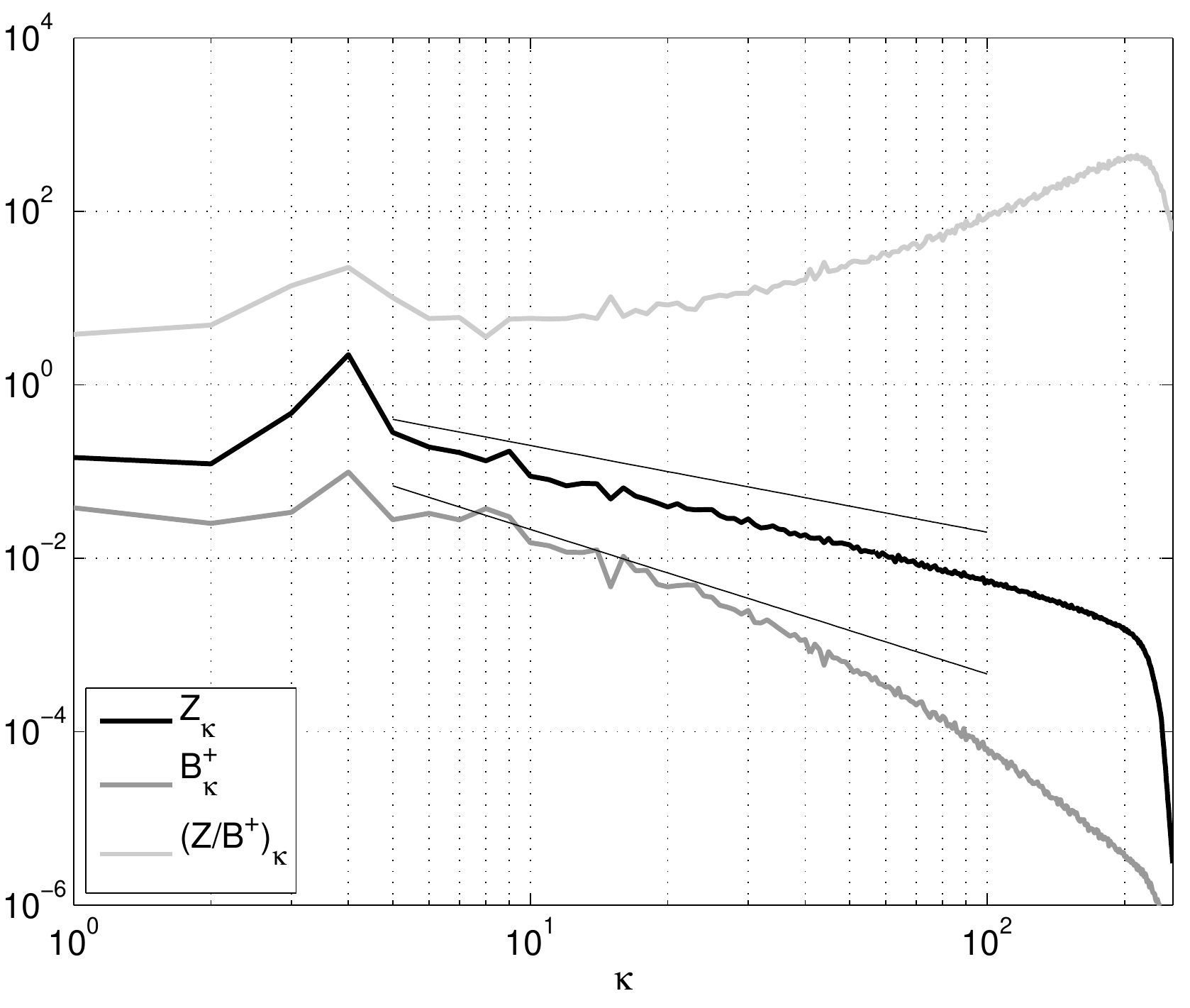}
\caption{Enstrophy $Z_\kappa$ and surface buoyancy variance
  $B^+_\kappa$ as functions of wavenumber $\kappa$ for the Ocean
  simulation (lines with slopes -1 and -5/3 are included for
  reference).  The ratio $Z_\kappa/B^+_\kappa$, also shown, can be used to
  guide the choice of the weight $\alpha_+$ for an effective
  projection basis.}
\label{fig:ZBratio}
\end{center}
\end{figure}

The limit $\alpha_- \to \infty$ would seem a natural choice of
generalized basis for typical ocean conditions takes because of the
relative lack of buoyancy activity at the bottom. Regarding
$\alpha_+$, an optimal value can in principle be chosen by inspecting
the spectra for a range of values or by using a diagnostic such as
that of Fig.~\ref{maxEmode}. However, some simpler rules of thumb
would be desirable. Intuitively, one might expect that the optimal values of
$\alpha_\pm$ are those that balance the contributions of the enstrophy
$Z_\kappa$ and of the surface-buoyancy variance $B^+_\kappa$ in the
generalized enstrophy $P_\kappa = Z_\kappa + \alpha_+ B^+_\kappa$.  Some support
for this intuition is provided by Fig.~\ref{fig:ZBratio} which shows
$Z_\kappa$, $B_\kappa$ and their ratio as a function of $\kappa$ for
the ocean simulation.  The figure shows a ratio $Z_\kappa/B^+_\kappa$
that is around $5$ for a broad range of $\kappa$, roughly consistent
with the value $\alpha_+=2$ indicated by Fig.~\ref{maxEmode}. There
is, however, a peak around $\kappa = 4$ and a substantial increase for
$\kappa \gtrsim 20$, which suggest that better results could be
obtained by allowing $\alpha_+$ to depend on $\kappa$. We have not
explored this intriguing possibility here.

As an alternative to the ratio $Z_\kappa/B^\pm_\kappa$, it would be
useful to relate more directly the value of the weights $\alpha_\pm$
most appropriate to project a flow on the large-scale characteristics
of the flow. Since for flows driven by instabilities, $Z_\kappa$ and
$B^\pm_\kappa$ are related to the large-scale PV and surface-buoyancy
gradients $Q_y$ and $B_y^\pm$, it is plausible that the ratio
$Q_y/B^\pm_y$ can be used as a guide for the choice of the weights.

%[speculate about setting $\alpha_\pm$ --- perhaps surface limit is
%generally right?  How much energy is captured by modes 3 and 4 for BC1
%with small-alpha modes?   Also discuss physical meaning of length
%scale associated with imaginary eigenvalues...  Also discuss BCI and
%speculate about choosing weights based on mean flow ratios $B_y/Q_y$.]

The advent of higher-resolution satellite observations,
expected when the Surface Water Ocean Topography satellite becomes
operational \citep{FuFerrari08}, will improve our understanding of upper-ocean
submesoscale dynamics only to the extent that we can connect surface
observations with the three-dimensional structure of the flow below
the surface.  The basis derived and demonstrated here may prove a
useful tool in this goal.

\clearpage

\begin{acknowledgment}
  KSS acknowledges the support of both NSF award OCE-0962054 and ONR
  award N00014-09-01-0633, and helpful conversations with Shane
  Keating and Xiao Xiao.  JV acknowledges the support of a Leverhulme
  Research Fellowship and the hospitality of the Courant Institute
  where this research was initiated.
\end{acknowledgment}

\clearpage

\begin{appendix}[A]
\section*{Derivation details} 
\label{app:deriv}

Here we prove a few relevant facts about the eigenvectors and
eigenvalues of \eqref{eval}.  First, we show that the operator
$\ca{E}$ is self-adjoint, e.g. $\< \bs{\xi}_m , \ca{E} \bs{\xi}_n
\>=\< \ca{E}\bs{\xi}_m , \bs{\xi}_n \>$.  Expanding the left-hand
side and integrating by parts, we find
\begin{align*}
\<\bs{\xi}_m , \ca{E}\bs{\xi}_n \> &=
\frac{1}{H}\int_{z^-}^{z^+}-\bar{\xi}_m \phi_n \, \mathrm{d}z 
+ \bar{\xi}_m^+ \phi_n(z^+) - \bar{\xi}_m^-\phi_n(z^-),\\
&= \frac{1}{H}\int_{z^-}^{z^+} -\phi_n\left(\frac{f^2}{N^2}\bar{\phi}_m'\right)' +\kappa^2\bar{\phi}_m \phi_n
\, \mathrm{d}z \\
&\qquad + \frac{f^2}{HN^2(z^+)}\bar{\phi}_m'(z^+) \phi_n(z^+) 
- \frac{f^2}{HN^2(z^-)}\bar{\phi}_m'(z^-)\phi_n(z^-),\\
&=\frac{1}{H}\int_{z^-}^{z^+}\frac{f^2}{N^2}\phi_n'\bar{\phi}_m' + \kappa^2\bar{\phi}_m \phi_n 
\, \mathrm{d}z,\\
%
%&=\frac{1}{H}\int_{z^-}^{z^+}\kappa^2\bar{\phi}_m \phi_n
%-\bar{\phi}_m\left(s\phi_n'\right)'~\mathrm{d}z 
%+ \frac{s(z^+)}{H}\bar{\phi}_m(z^+) \phi_n'(z^+) 
%- \frac{s(z^-)}{H}\bar{\phi}_m(z^-)\phi_n'(z^-),\\
&= \< \ca{E}\bs{\xi}_m , \bs{\xi}_n \>
\end{align*}
since the expression on the penultimate line is clearly symmetric. The self-adjointness of $\mathcal{P}$ as well as the positive definiteness is obvious.

To establish the completeness of the basis of the eigenvector $\bs{\xi}_n$, we rewrite the eigenvalue problem in the standard form $\mathcal{A} \bs{\xi}_n = \mu_n^{-2} \bs{\xi}_n$, where $\mathcal{A}=\mathcal{P}^{-1} \mathcal{E}$ is positive definite and self-adjoint. This operator is compact when acting on the Hilbert space of vectors $\bs{Q}$ with bounded norm $\< \bs{Q},\bs{Q}\>$. 
This is because it is essentially an integral operator with continuous
kernel --- the Green's function of the operator $(s \phi')'-\kappa^2 \phi$ \citep[e.g.][section 4.8]{Denbath-Miku}. The Hilbert-Schmidt theorem \citep[section 4.10]{Denbath-Miku} then applies to guarantee that every vector $\bs{Q}$ has a unique convergent expansion in terms of the $\bs{\xi}_n$.
\end{appendix}

\clearpage

\begin{appendix}[B]
\section*{Discrete eigenvalue problem and numerical computation of modes}
\label{app:numerics}

Here we construct the discrete version of the eigenvalue
problem. Assuming a constant discrete coordinate $z_j$ on $J$ grid
points, with $z_1=0$ at the top, $z_J=-H$ at the bottom, and a
constant finite difference $\Delta z = z_j-z_{j+1}$, the mean
stratification is $N_0^2 = (g/\rho_0)\ol{\Delta\rho}/\Delta z$,
where $\ol{\Delta\rho} = \rho_J-\rho_1$ is the average background
density jump between levels, $\rho_j = \rho(z_j)$ is the background
density, and $\rho_0$ is the average density.  The parameter
$s = N_0^2/N^2$ is discretized as $s_{j} = s(z_{j+1/2}) \equiv
\ol{\Delta\rho}/(\rho_{j+1}-\rho_j)$, thus $s_j$ is offset by a
half space from $\rho_j$.  In this discretization, the SBs and PV are
\begin{align*}
b^+ = \frac{f^2}{N_0^2H}~s\psi'|_{z=0} ~\longrightarrow ~
&L_D^{-2}~\frac{s_1}{\delta}(\psi_1-\psi_2)\\
b^-  = \frac{f^2}{N_0^2H}~s\psi'|_{z=-1} ~\longrightarrow~ 
&L_D^{-2}~\frac{s_{J-1}}{\delta}(\psi_{J-1}-\psi_J)\\
q = \left(\frac{f^2}{N_0^2}s\psi'\right)' -\kappa^2\psi ~\longrightarrow ~
&L_D^{-2}~\frac{1}{\delta^2}\left[s_{j-1}\psi_{j-1} - (s_{j-1}+s_j
  )\psi_j + s_j\psi_{j+1}\right] -\kappa^2 \psi_j,
\end{align*}
where $\delta \equiv \Delta z/H$ and $L_D \equiv N_0H/f$.  Nondimensionalizing
$\kappa \mapsto [L_D^{-1}]~\kappa$, $\psi \mapsto [L_D^2T^{-1}]
~\psi$ and $(q,b^\pm) \mapsto [T^{-1}]~ (q,b^\pm)$ (for some timescale $T$), the
discrete PV/SBs and streamfunction are related as 
\[
\bs{Q} =\mat{A}\bs{\psi}, 
\]
where
\begin{equation}\label{Adef}
\mat{A} = \frac{1}{\delta^2}\begin{pmatrix}
\delta s_1 & -\delta s_1& 0 & \hdotsfor{3} &0\\
s_1 & -(s_1+s_2+\delta^2\kappa^2) & s_2 & 0 &   \hdotsfor{2} & 0\\
 \\
 & &  \hdotsfor{3} & \\
 \\
0 & \hdotsfor{2} & 0 & s_{J-2} &
-(s_{J-2}+s_{J-1}+\delta^2\kappa^2) & s_{J-1}\\
 0 & \hdotsfor{3} & 0 & \delta s_{J-1} &
-\delta s_{J-1}
\end{pmatrix}.
\end{equation}
Defining the operators
\begin{equation}\label{FBdef}
\mat{B} = \begin{pmatrix}
1 & 0 &  \hdotsfor{2} & 0\\
0 & \delta &  \hdotsfor{2} & 0\\
& & & \\
0 &  \hdotsfor{2} & \delta & 0\\
 0 & \hdotsfor{2} & 0 & 1
\end{pmatrix}
~~~\text{and}~~~
\mat{F} = \begin{pmatrix}
1 & 0 &  \hdotsfor{2} & 0\\
0 & -1 &  \hdotsfor{2} & 0\\
& & & \\
& & & \\
0 & \hdotsfor{2} & 0 & -1
\end{pmatrix},
\end{equation}
one sees that $\mat{B}$ plays the part of the inner product,
e.g. $\<\bs{\xi}_1,\bs{\xi}_2\> \rightarrow
\bs{\xi}_1^T\mat{B}\bs{\xi}_2$ and $\mat{F}$ accomplishes the awkard
sign changes in the definition of the operator $\ca{E}$.  The energy
in wavenumber $\kappa$ is 
\[
E_\kappa =\frac{\delta}{2}\left[\sum_{j=1}^{J-1}s_j\left|\frac{\psi_j-\psi_{j-1}}{\delta}\right|^2
+\kappa^2\sum_{j=2}^{J-1}|\psi_j|^2\right]= \frac{1}{2}\bs{\psi}^*\mat{F}\mat{B}\mat{A}\bs{\psi}.
\]
For consistency with the theoretical development in section 2, we may
also write the energy in terms of the vector $\bs{Q} =
\mat{A}\bs{\psi}$,
\[
E_\kappa = \frac{1}{2}\bs{Q}^*\mat{B}\mat{F}\mat{A}^{-1}\bs{Q}
=  \frac{1}{2}\bs{Q}^*\mat{B}\ca{E}\bs{Q}
\]
where the symmetry of $\mat{F}$ and $\mat{B}$ were used, and $\ca{E}
\equiv \mat{F}\mat{A^{-1}}$ is defined to make the discrete version of
the energy operator defined in \eqref{EPdef} perfectly clear.

Similarly, the generalized enstrophy in wavenumber $\kappa$ is 
\[
P_\kappa = \frac{1}{2}\bs{Q}^*\mat{B}\ca{P}\bs{Q}
\]
where we define
\[
\ca{P} = \begin{pmatrix}
\alpha_+ & 0 &  \hdotsfor{2} & 0\\
0 & 1 &  \hdotsfor{2} & 0\\
& & & \\
0 &  \hdotsfor{2} & 1 & 0\\
 0 & \hdotsfor{2} & 0 & \alpha_-
\end{pmatrix}
\]
to make clear the analogy with the generalized enstrophy operator
defined in \eqref{EPdef}.
 
Now note that $\mat{B}\ca{E}$ and $\mat{B}\ca{P}$ are both symmetric
(the former can be verified by checking that $\mat{F}\mat{B}\mat{A}$ is
symmetric), so we can simultaneously diagonalize the two quadratic
forms $E_\kappa$ and $P_\kappa$ by solving the generalized eigenvalue
problem $\mat{B}\ca{P}\bs{\xi}_j = \mu_j^2\mat{B}\ca{E}\bs{\xi}_j$
or, in matrix form
\[
(\mat{B}\ca{P})\mat{X} = (\mat{B}\ca{E})\mat{X}\mat{M}^2
\]
where $\mat{X}$ is the matrix with columns $\bs{\xi}_j$ and
$\mat{M}^2$ has $\mu_j^2$ along is its diagonal and zeros elsewhere.
Solutions to this generalized eigenvalue problem obey the
orthogonality relations
\begin{equation}\label{xorthog}
\mat{X}^\top\mat{B}\ca{E}\mat{X} = \mat{I} ~~~\text{and}~~~
\mat{X}^\top\mat{B}\ca{P}\mat{X} = \mat{M}^2,
\end{equation}
which are analogous to \eqref{ortho} and \eqref{xiPxiortho}, respectively.

In practice, it is more convenient to define a streamfunction
eigenfunction $\bs{\phi}$ such that $\mat{A}\bs{\phi} = \bs{\xi}$,
so that the generalized eigenvalue problem can be rewritten as
$\mat{F}\ca{P}\mat{A}\bs{\phi}_j = \mu_j^2\bs{\phi}_j$, or in matrix
form
\begin{equation}\label{phigenval}
\mat{F}\ca{P}\mat{A}\mat{\Phi} = \mat{\Phi}\mat{M}^2
\end{equation}
where $\mat{\Phi}$ has $\bs{\phi}_j$ as its columns.  In this case,
the orthogonality relations become
\begin{equation}\label{porthog}
\mat{\Phi}^\top\mat{F}\mat{B}\mat{A}\mat{\Phi} = \mat{I} ~~~\text{and}~~~
\mat{\Phi}^\top\ca{P}\mat{B}\mat{A}^2\mat{\Phi} = \mat{M}^2,
\end{equation}
where we've used the fact that $\mat{F}^2 =\mat{I}$.
Finally, writing \eqref{phigenval} as
$\mat{\Phi}^{-1}(\mat{A}^{-1}\ca{P}^{-1}\mat{F})\mat{\Phi} = \mat{M}^2$
and using the first relation in \eqref{porthog}, we have the
equivalent of \eqref{PExiortho},
\begin{equation}\label{phiorthgo}
\mat{\Phi}^{-1}\mat{B}\ca{P}^{-1}\mat{\Phi} = \mat{M}^{-2}
\end{equation}

The expansion in the basis of eigenvectors $\vec{\phi}_n$ of discrete
data is readily expressed in terms of the matrix
$\mat{\Phi}$. Denoting by $\vec{\psi}$ the column vector of the
streamfunction data (Fourier transformed in the horizontal)
$\psi(z_j)$, the expansion reads
\begin{equation} \label{disproj}
\vec{\psi} = \mat{\Phi} \vec{a},
\end{equation}
where $\vec{a}=(a_1,\ldots,a_J)^\top$ is the column vector of
the mode amplitudes.  These amplitudes are obtained from the data
using the relation
\[
\vec{a} = \mat{\Phi}^\top \mat{F}\mat{B} \mat{A} \vec{\psi},
\]
which is deduced from \eqref{porthog} and \eqref{disproj}. The total
energy at a given wavenumber $\kappa$,
\[
E_\kappa = \frac{1}{2} \vec{\psi}^*\mat{F}\mat{B}\mat{A}\vec{\psi} =
\frac{1}{2}|\vec{a}|^2, 
\]
where $*$ denotes the complex (conjugate) transpose, is clearly the
sum of the individual contributions $|a_n|^2/2$ of each
mode. Similarly, the generalized enstrophy,
\[
P_\kappa = \frac{1}{2}\bs{Q}^*\mat{B}\ca{P}\bs{Q} 
=\frac{1}{2} \vec{\psi}^*\ca{P}\mat{B}\mat{A}^2 \vec{\psi} 
=\frac{1}{2} \vec{a}^* \mat{M}^2 \vec{a},
\]
is the sum of the contributions $\mu_n^2 |a_n|^2/2$. 
\end{appendix}

\clearpage

\bibliographystyle{ametsoc}
\bibliography{lit}

\clearpage

\clearpage

\clearpage

\clearpage

\clearpage

\clearpage

\clearpage

\clearpage

\end{document}